\pdfoutput=1

\documentclass[11pt]{article}

\usepackage[dvipsnames]{xcolor}

\usepackage[final]{acl}

\usepackage{times}
\usepackage{latexsym}

\usepackage[T1]{fontenc}

\usepackage[utf8]{inputenc}

\usepackage{microtype}

\usepackage{inconsolata}

\usepackage{float}
\usepackage{graphicx}
\usepackage{placeins}
\usepackage{paralist}
\usepackage{amsmath}
\usepackage{amssymb}
\usepackage{booktabs}
\usepackage{tcolorbox}
\usepackage{pgfplots}
\usepackage{pythonhighlight}

\PassOptionsToPackage{table}{xcolor}
\usepgfplotslibrary{groupplots}
\pgfplotsset{compat=1.18}

\tcbset{
  colframe=blue!20,
  colback=blue!10,
  coltitle=black,
  fonttitle=\bfseries
}

\title{CodeArena: A Collective Evaluation Platform for LLM Code Generation}

\author{
    \textbf{Mingzhe Du\textsuperscript{1,2}},\quad
    \textbf{Luu Anh Tuan\textsuperscript{1}\thanks{Corresponding Author.}},\quad
    \textbf{Bin Ji\textsuperscript{2}}, \quad
    \textbf{Xiaobao Wu\textsuperscript{1}},\quad
    \textbf{Dong Huang\textsuperscript{2}}, \\
    \textbf{Yuhao Qing\textsuperscript{3}},\quad
    \textbf{Terry Yue Zhuo\textsuperscript{4}},\quad
    \textbf{Qian Liu\textsuperscript{5}},\quad
    \textbf{See-Kiong Ng\textsuperscript{2}}
    \\
    \textsuperscript{1}Nanyang Technological University \quad \textsuperscript{2}National University of Singapore \\
    \textsuperscript{3}The University of Hong Kong \quad \textsuperscript{4}Monash University \quad \textsuperscript{5}TikTok \\
    \texttt{\{mingzhe001,anhtuan.luu,xiaobao.wu\}}@ntu.edu.sg, \texttt{qyhh}@connect.hku.hk, \\
    \texttt{\{jibin,dhuang,seekiong\}}@nus.edu.sg, \texttt{terry.zhuo}@monash.edu, \texttt{qian.liu}@tiktok.com
}

\begin{document}
\maketitle

\begin{abstract}
Large Language Models~(LLMs) have reshaped code generation by synergizing their exceptional comprehension of natural language and programming syntax, thereby substantially boosting developer productivity. 
While these advancements have spurred many efforts to quantitatively assess LLM coding abilities, persistent issues like benchmark leakage, data dissipation, and limited system accessibility hinder timely and accurate evaluations.
To address these limitations, we introduce \texttt{CodeArena}\footnote{Website: \url{https://codearena.online}}$^,$\footnote{Demo Video: \url{https://youtu.be/yqF9Cdrh3ss}}, an online evaluation framework tailored for LLM code generation. 
The key innovation is a collective evaluation mechanism, which dynamically recalibrates individual model scores based on the holistic performance of all participating models, mitigating score biases caused by widespread benchmark leakage. 
In addition, \texttt{CodeArena} ensures open access to all submitted solutions and test cases and provides automation-friendly APIs to streamline the code evaluation workflow. 
Our main contributions are: (1)~a collective evaluation system for unbiased assessment, (2)~a public repository of solutions and test cases, and (3) automation-ready APIs for seamless integration.
\end{abstract}

\section{Introduction}

\begin{figure}[t]
    \centering
    \includegraphics[width=0.99\linewidth]{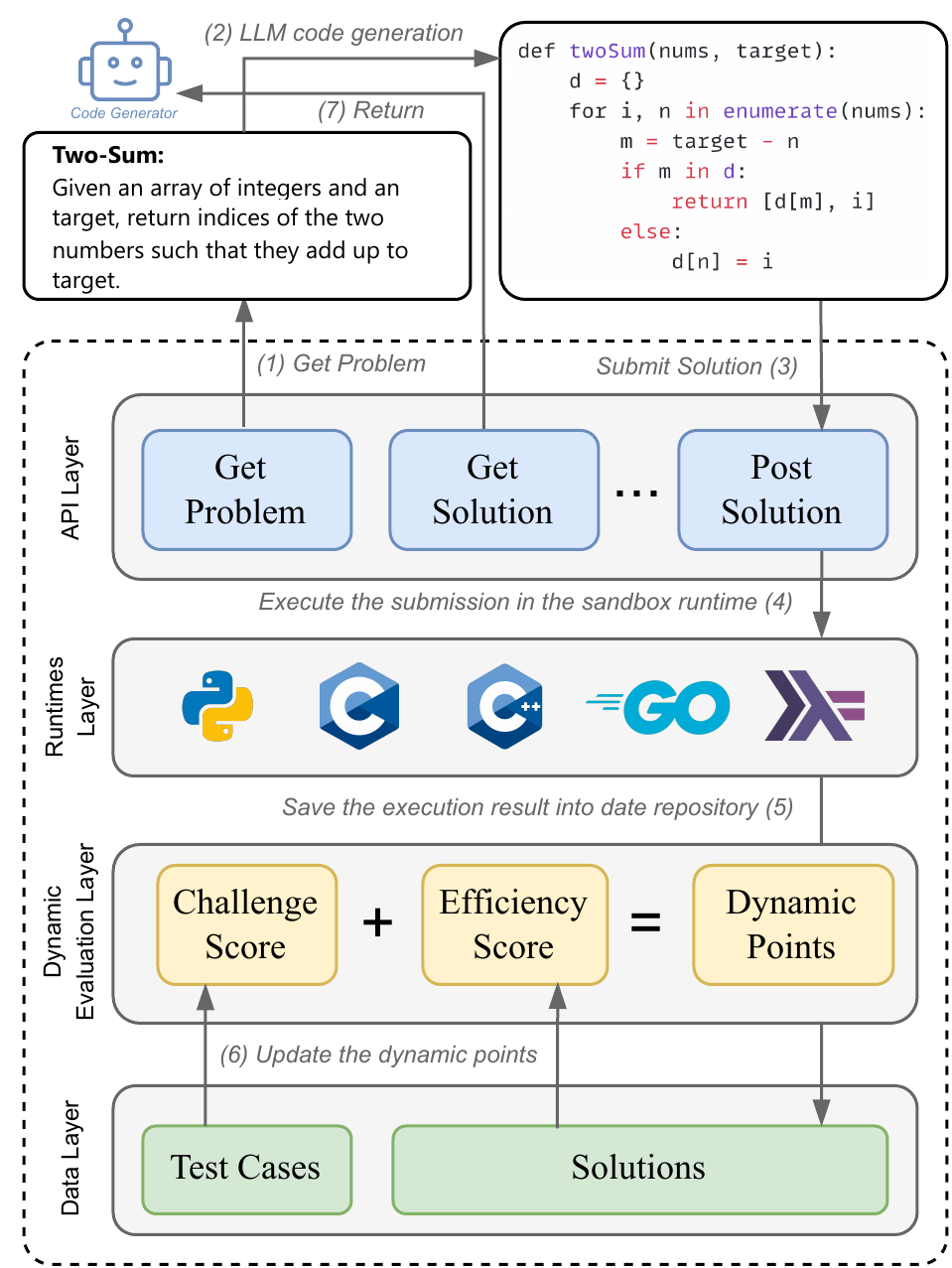}
    \caption{
        The \texttt{CodeArena} framework allows users to interact with the system through APIs. 
        The depicted workflow shows the code submission process.
    }
    \label{fig:codearena_pipeline}
\end{figure}

Leveraging the exceptional language comprehension and generation capabilities of large language models (LLMs), automatic code generation has significantly transformed the landscape of software development~\cite{lozhkov2024starcoder, roziere2023code, zhu2024deepseek,huang2024agentcoder}. By interpreting natural language instructions, LLMs can now directly generate code, introducing new efficiencies and possibilities in the software development process. 
To evaluate the performance of LLMs in code generation, various benchmarks have emerged that assess the generated code from multiple perspectives. For instance, HumanEval~\cite{chen2021evaluating} and its successors~\cite{evalplus,zhuo2024bigcodebench} are widely used to assess the functional correctness of LLM-generated code. 
Beyond the functional correctness, Mercury~\cite{du2024mercury} and EffiBench~\cite{huang2024effibench} assess the efficiency of LLM-generated code, while CyberSecEval~\cite{bhatt2024cyberseceval} quantifies LLM security risks.
Furthermore, online judge~(OJ) platforms, such as LeetCode~\cite{leetcode2024} and CodeForces~\cite{codeforces2024}, offer online code assessment services, enabling code evaluation against predefined test cases.

Although existing evaluation approaches have achieved great success, {they have three} limitations:

\textbf{(1) Benchmark Contamination.} 
Leakage of benchmark data into LLM training datasets can result in contamination, causing LLMs to perform abnormally on benchmarks~\cite{jain2024livecodebench,wu2024akew,wu2024antileak}. Regularly importing new problems into the evaluation can alleviate this issue. However, given the static and offline nature of most code evaluation benchmarks, it is hard to distribute the up-to-date benchmark to each LLM and dynamically get the real-time performance evaluation. Moreover, current benchmarks for LLM code generation predominantly evaluate individual models in isolation, neglecting holistic factors. 
For instance, the difficulty of most problems is defined subjectively by human curators, which may not accurately represent the challenge posed to LLMs.

\textbf{(2) Data Dissipation.} 
Most existing benchmarks merely record the final metrics, while discarding the generated code solutions. Similarly, many online platforms do not make user-submitted solutions publicly accessible~\cite{leetcode2024, DMOJ_2024}. However, such solution data is crucial for advancing LLM code generation research. For example, to evaluate the execution efficiency of LLM-generated code, the Mercury benchmark requires a sufficient amount of solutions to analyze the distribution of execution times~\cite{du2024mercury}. Additionally, fine-tuning the code generation capabilities of LLMs necessitates a substantial dataset of $\langle \text{problem}, \text{solution} \rangle$ pairs as well.

\textbf{(3) System Accessibility.} 
Current code generation benchmarks employ disparate evaluation protocols, often necessitating local execution or manual submission to leaderboards~\cite{zhuo2024bigcodebench, chen2021evaluating, evalplus}. 
This complexity not only complicates model evaluation but also makes it difficult to keep pace with the rapid LLM advancements. Although OJ platforms, such as Leetcode and DMOJ, offer unified online code evaluation services, they lack automation-friendly application programming interfaces~(APIs) for submitting LLM-generated code. Consequently, researchers are compelled to use automation testing tools like Selenium to submit code to these platforms~\cite{du2024mercury, huang2024effibench}, impeding rapid model evaluation. 

To address these challenges, this paper introduces \texttt{CodeArena}, an online evaluation framework tailored for LLM code generation. 
Regarding the data contamination issue, \texttt{CodeArena} proposes a novel dynamic scoring mechanism instead of merely relying on the integration of new problems. The newly introduced metric, \texttt{Dynamic Point}, assigns rewards to each accepted solution in a way that ensures even widespread leakage of an evaluation problem has minimal impact on the benchmark results. This approach effectively mitigates the influence of data contamination.
In addition to serving as an assessment platform, \texttt{CodeArena} functions as a solution repository. Rather than discarding submitted solutions after evaluation, \texttt{CodeArena} systematically records them and makes them publicly accessible.
Moreover, to facilitate seamless user interaction, \texttt{CodeArena} offers a suite of automation-friendly APIs. 

The main contributions are summarized as follows:
\textbf{1) Dynamic Evaluation.} We introduce \texttt{CodeArena}, an OJ framework that periodically integrates novel coding tasks to ensure they remain uncontaminated, and dynamically adjusts scoring metrics to effectively evaluate the code generation capabilities of LLMs.
\textbf{2) Open Data Repository.} All \emph{solutions} and \emph{test cases} are publicly accessible, promoting an open-source environment conducive to analyzing and improving LLM code generation.
\textbf{3) Automation-friendly APIs.} We provide APIs designed to streamline the automated code evaluation process, facilitating efficient user interaction.

\section{Related Work}
\subsection{Code Assessment Platforms}
LeetCode~\cite{leetcode2024} is a prominent online coding platform that offers an extensive array of problems across diverse domains such as algorithms, data structures, databases, and system design. The platform provides instant feedback and detailed analysis of code performance, enabling users to iteratively refine their solutions. Similarly, CodeForces~\cite{codeforces2024} is another well-regarded competitive platform, renowned for its regular contests and vast, crowd-sourced collection of programming problems. Unlike these closed-source platforms, DMOJ~\cite{DMOJ_2024} provides an open-source OJ framework, which includes the front-end user interface, runtime environments, and API endpoints. Despite offering plentiful coding evaluation resources, these platforms are not designed for automated LLM submissions. CodeArena bridges this gap by integrating these resources and providing automation-friendly APIs specifically for evaluating LLM-generated code.

\subsection{Code Generation Benchmarks}
Most code generation benchmarks adopt a fuzzing methodology~\cite{zhu2022fuzzing, hendrycksapps2021,huang2024effibench,qing2025effibenchx,ouyang2025dsbench,dai2024mhpp,huang2025bias}, where predefined test cases are executed on the generated code, and the outputs are compared to expected results. For example, HumanEval~\cite{chen2021evaluating} comprises 164 handcrafted programming problems and emphasizes the functional correctness of generated code. BigCodeBench~\cite{zhuo2024bigcodebench} extends this evaluation framework by including more complex instructions and diverse function calls, thus testing the true programming capabilities of LLMs in realistic scenarios. LiveCodeBench~\cite{jain2024livecodebench} takes a step further by continuously updating its problem set, ensuring contamination-free evaluations. Additionally, recognizing the gap in evaluating computational efficiency, Mercury~\cite{du2024mercury} introduces an efficiency-centric benchmark that considers the holistic runtime distribution, thereby assessing both the correctness and efficiency simultaneously.

\section{CodeArena}
\begin{figure*}[ht]
    \centering
    \includegraphics[width=\linewidth]{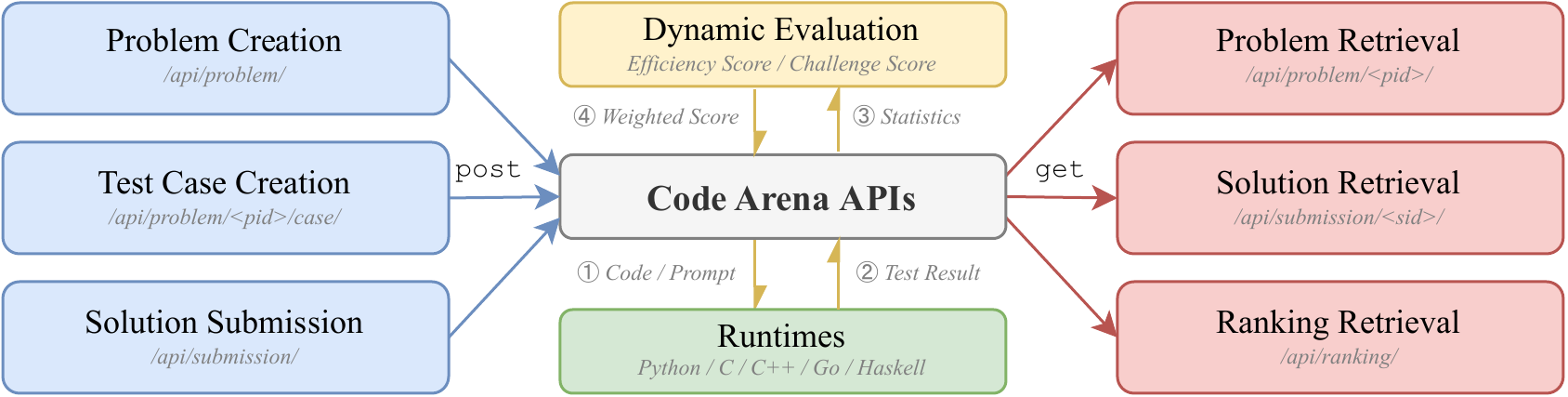}
    \caption{Overview of \texttt{CodeArena}. The \emph{\textcolor{LimeGreen}{Green}} component provides runtime environments for programming languages, capable of accepting either generated code or model prompt as the input, and outputting test results. The \emph{\textcolor{Dandelion}{Yellow}} component is the dynamic evaluation unit, updating the LLM weighted ranking score based on each submission result. The \emph{\textcolor{Blue}{Blue}} and \emph{\textcolor{Maroon}{Maroon}} components are RESTful API \emph{POST~($\lhd$)} and \emph{GET~($\rhd$)} calls, respectively.}
    \label{fig:overview}
\end{figure*}

As depicted in Figure~\ref{fig:codearena_pipeline}, \texttt{CodeArena} is an online code evaluation platform built upon an open-source OJ framework DMOJ~\cite{DMOJ_2024}. The platform is structured into four distinct layers:
\emph{The API Layer} provides a set of APIs to facilitate user interactions. \emph{The Runtimes Layer} offers a standardized environment for code execution and evaluation. \emph{The Dynamic Evaluation Layer} processes execution results from \emph{the Runtimes Layer} and dynamically updates ranking scores after each submission. Finally, \emph{the Data Layer} stores problems, test cases, and solutions. 
In this section, we will delve into the \texttt{CodeArena} framework breakdown~(Section~\ref{sec:framework_breakdown}) and the detailed workflows~(Section~\ref{sec:workflows}).

\subsection{Framework Breakdown}
\label{sec:framework_breakdown}
\paragraph{API Layer.}
While existing OJ platforms like LeetCode and DMOJ offer online code assessment services, a significant limitation for LLM researchers is the lack of automation-friendly APIs. Researchers are compelled to harness automation testing tools to submit LLM-generated code, which can be cumbersome. To address this, \texttt{CodeArena} provides an automation-friendly interface via a set of REST APIs~\cite{rodriguez2016rest} and a dedicated Python library, \emph{codearena}\footnote{\url{https://pypi.org/project/codearena/}}, enabling streamlined code submission to our platform. As illustrated in Figure~\ref{fig:overview}, \texttt{CodeArena} offers endpoints for \texttt{Authentication}, \texttt{Problem}, and \texttt{Ranking} utilizing standard RESTful API methods \emph{POST}~($\lhd$) and \emph{GET}~($\rhd$)~\cite{richardson2008restful}:

\noindent \textbf{$\lhd$~Authentication}~(\url{/api/authentication/}): To ensure secure submissions and data retrieval, we require all registered users to generate an \emph{API Token} to access \texttt{CodeArena}. The \emph{API Token} can be revoked and regenerated as necessary.

\noindent \textbf{$\lhd$~Problem Creation}~(\url{/api/problem/}): We encourage the submission of new problems to diversify the problem set. Authorized benchmark curators can manage and distribute new problems via this API. For instance, LiveCodeBench~\cite{jain2024livecodebench} can regularly submit new problems to CodeArena, and the platform will automatically test and update the ranking of all code generator users with these new problems.

\noindent \textbf{$\lhd$~Test Case Creation}~(\url{/api/problem/<pid>/case}): High-quality test case collection is challenging \cite{huang2025measuring}, as most OJ platforms do not release the test cases used for problem assessment. To solve this issue, \citet{du2024mercury}, \citet{huang2024effibench}, and \citet{huang2025swiftcoder} utilize GPT models \cite{achiam2023gpt} to write test case generators. In our work, we follow the same way to gather initial test cases for each problem and encourage users to upload their own test cases. Here, $\langle pid \rangle$ denotes the specific problem ID.

\noindent \textbf{$\lhd$~Solution Submission}~(\url{/api/submission}): \emph{Code Generator} users can submit their generated code for a specific $\langle pid \rangle$ problem via this API. \texttt{CodeArena} executes the submitted code in a sandbox and returns a $submission\_id$ to the user, which is then used with \emph{Solution Retrieval} to retrieve the detailed execution status.

\noindent \textbf{$\rhd$~Problem Retrieval}~(\url{/api/problem/}): This API has two variants: \url{/api/problem/} lists all problems with their corresponding IDs, whereas \url{/api/problem/<pid>/} provides detailed information, such as problem descriptions and acceptance statistics, for a specific problem $\langle pid \rangle$.

\noindent \textbf{$\rhd$~Solution Retrieval}~(\url{/api/submission/}): Similar to problem retrieval, solution retrieval has two variants: \url{/api/submission/} lists all submissions, and \url{/api/submission/<sid>} provides runtime information for a specific submission $\langle sid \rangle$.

\noindent \textbf{$\rhd$~Ranking Retrieval}~(\url{/api/ranking}): This endpoint returns real-time ranking results in JSON format, identical to those shown on \url{https://codearena.online/users/}.

\paragraph{Runtimes Layer.}
To ensure the stable and secure execution of code submissions, CodeArena operates within an isolated sandbox runtime environment\footnote{\url{https://github.com/Elfsong/Monolith}}. This environment currently supports multiple programming languages, including \textit{Python 3, C, C++, Go, and Haskell}, while holding the flexibility to integrate additional languages as needed. The runtime system reports both running time and memory overhead for each submission, and it raises exceptions and provides detailed error information if a code submission fails to execute properly.

The \texttt{CodeArena} runtime environment accommodates two types of inputs: 
\textbf{Code runtime} directly accepts and executes code submitted by a code generator.
\textbf{Prompt runtime} is designed for interactions with LLMs. Instead of submitting raw code, code generators provide a model prompt. The runtime then uses this prompt to invoke the appropriate LLM locally, and the generated code is subsequently executed within the sandbox. 

\paragraph{Dynamic Evaluation Layer.}

\begin{figure*}[ht]
    \centering
    \includegraphics[width=0.9\textwidth]{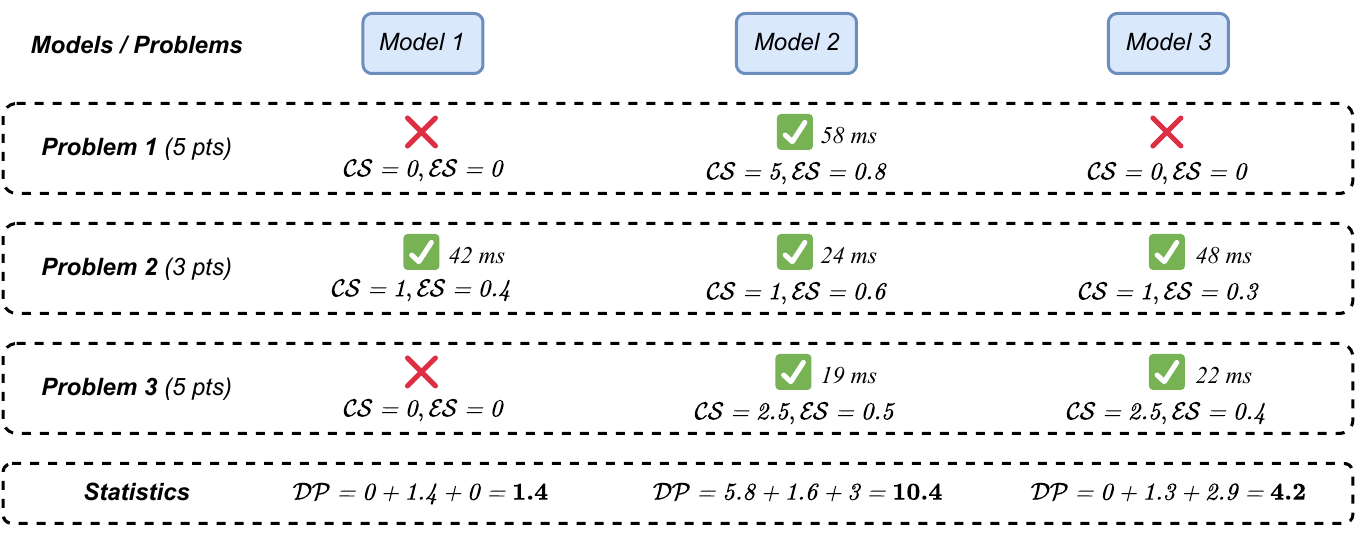}
    \caption{Example of \texttt{Dynamic Point}~($\mathcal{DP}$) calculation. Each individual model score is influenced by the overall system performance. $\mathcal{CS}$ and $\mathcal{ES}$ are counted only when the model passes (\checkmark) all test cases. Runtimes and $\mathcal{ES}$ values are illustrative; in practice, $\mathcal{ES}$ is the runtime percentile among all accepted solutions to the problem on the platform, not only among the three models shown.}
    \label{fig:evaluation_example}
\end{figure*}

Solving problems with varying difficulty levels should contribute accordingly to the ranking score. However, the difficulty of most benchmark problems is determined subjectively by data curators, which may not accurately represent the challenges posed to LLMs~\cite{du2024evaluating}. 
As illustrated in Figure~\ref{fig:difficulty_distribution}, the acceptance rate~($\mathcal{AC}$) does not show significant variation across difficulty levels. To rectify this discrepancy, we propose the \emph{Challenge Score} ($\mathcal{CS}$):
\begingroup\predisplaypenalty=0 
\begin{equation}
    \mathcal{CS}_i = \frac{\mathcal{BPS}_i}{S^{solved}_i} = \frac{\mathcal{BPS}_i}{\mathcal{AC}_i \times S^{total}_i},
\end{equation}
\endgroup%
where $\mathcal{BPS}_i$ represents the basic problem score of the $i$-$th$ problem, $S^{solved}_i$ is the number of users who solved it, and $\mathcal{AC}_i = S^{solved}_i / S^{total}_i$ denotes the acceptance rate. Essentially, all users who solve the problem share $\mathcal{BPS}_i$ evenly, while those who fail receive $\mathcal{CS}_i = 0$. Solving an easy problem that most users can solve yields a minimal bonus, whereas solving a challenging problem earns a higher $\mathcal{CS}_i$. 
For instance, consider a problem worth 5 points: if only one LLM successfully solves it, that model receives the full 5 points. However, if all LLMs solve the problem, indicating either widespread leakage or a lack of discriminatory difficulty, the 5 points are distributed evenly among them. This ensures that leaked or overly simplistic problems have minimal influence on the overall leaderboard, effectively mitigating the risk of data contamination.

Moreover, \texttt{CodeArena} also considers the Efficiency Score~($\mathcal{ES}_i$) of the generated code by calculating the runtime percentile of the current solution~($s_c^{rt}$) relative to the runtime of other solutions~($s_j^{rt}$):
\begin{equation}
    \mathcal{ES}_i = \frac{\Vert {s_j \mid s_c^{rt} \leq s_j^{rt}, s_j \in S_i^{solved}}\Vert}{\Vert{S_i^{solved}}\Vert}.
\end{equation}

Therefore, the final \texttt{Dynamic Point}~($\mathcal{DP}$) for each user is given by:
\begin{equation}
    \mathcal{DP} = \sum_{i = 1}^{N} (\mathcal{CS}_i + \mathcal{ES}_i),
\end{equation}
where $N$ is the number of problems. We record the \texttt{Dynamic Point} ranking regularly to observe the performance trending of each user. Additionally, this layer supports customized metrics.

\paragraph{Data Layer.}
In addition to evaluating code generation capabilities, CodeArena is envisioned as a comprehensive open-source data platform. Its data layer is structured to store rich metadata for each problem, accompanied by a diverse collection of solutions with detailed execution overhead metrics. This robust dataset serves as a foundation for analyzing model performance trends and fostering advancements in code generation LLMs.

\subsection{Workflows}
\label{sec:workflows}
In this section, we outline the workflow for each \texttt{CodeArena} user group: \texttt{Benchmark Curators} (\textit{Problem Collection, Quality Control, Test Case Generation}), \texttt{Code Generators} (Code Submission), and \texttt{Data Readers}. Each user group is assigned specific tasks and granted distinct system permissions. A detailed definition of these user groups is provided in Appendix~\ref{sec:user_groups}.

\paragraph{Problem Collection.}
\label{sec:problem_collection}
To diversify our problem set and prevent benchmark leakage, we developed a workflow for \texttt{Benchmark Curators}. This workflow integrates existing code evaluation datasets, such as Mercury~\cite{du2024mercury} and APPS~\cite{hendrycksapps2021}, through dedicated scripts and can easily incorporate other benchmarks with structured problem descriptions and test cases. For online coding platforms, we primarily collect source problems from weekly contests on CodeForces\footnote{\url{https://codeforces.com/}} and LeetCode\footnote{\url{https://leetcode.com/problems/}}. 

\paragraph{Quality Control.} 
\label{sec:quality_control}
CodeArena initially populates its benchmark set using the APPS~\cite{hendrycksapps2021} and Mercury~\cite{du2024mercury} datasets. To ensure the quality of each task, a rigorous screening process with multiple filters is applied:
\textbf{1) Publicity:} Only publicly accessible and free questions from online resources are selected; proprietary or commercial questions are excluded.
\textbf{2) Task Category:} As CodeArena focuses on evaluating \emph{algorithmic} code generation, only tasks of this nature are retained.
\textbf{3) Sufficient Oracle Solutions:} Tasks must possess more than 16 reference solutions. This criterion ensures we have enough solutions to validate the subsequent generated test case generators.
\textbf{4) Test Case Validity and Quantity:} A task is included if 100 valid test cases can be generated for it within a reasonable timeframe~(360s in our setting). The validity of each generated test case is confirmed by feeding it to all oracle solutions and verifying that all outputs are identical. 
\textbf{5) Line Coverage:} The generated test cases must achieve at least 60\% line coverage when executed against the oracle solutions~\cite{li2006code}. Only tasks meeting this coverage requirement are incorporated into the final benchmark set.

\paragraph{Test Case Generation.}
Since most online coding platforms do not disclose their test cases, we develop an automated test case generation workflow for \texttt{Benchmark Curators} to address this limitation. 
After regularly gathering coding problems, we employ \texttt{GPT-4o} to generate corresponding test case generators for each problem. For instance, consider the example problem: \textit{``Given an array of integers \textbf{nums} and an integer \texttt{target}, return indices of the two numbers such that they add up to target. You may assume that each input would have exactly one solution.''} 
For this problem, \texttt{GPT-4o} returns a test case generator as provided below. 

\begin{python}
from random import randint

def test_case_generation():
    n =  randint(2, 10**4)
    v1 = randint(-10**9, 10**9)
    v2 = randint(-10**9, 10**9)
    target = v1 + v2
    nums = [v1, v2]
    while len(nums) < n:
        v = randint(-10**9, 10**9)
        if (target - v not in nums):
            nums.append(v)
    return nums, target
\end{python}

Subsequently, we generate diverse test cases by randomly invoking the produced \texttt{test\_case\_generation} function.
As demonstrated in Mercury~\cite{du2024mercury}, LLM-generated test-case generators do not introduce bias towards specific LLMs. Additionally, to ensure the validity of each generated test case, we retain only those that yield consistent outputs across all canonical code solutions.

\paragraph{Code Submission.}
As shown in Figure~\ref{fig:codearena_pipeline}, the workflow for \texttt{Code Generators} comprises the following steps:
\textbf{1) Problem Retrieval.} A \texttt{Code Generator} initiates the workflow by calling the \url{/api/problem/} {Get API}, which retrieves the problem description.
\textbf{2) Code Generation.} Upon receiving the problem description, the \texttt{Code Generator} invokes the corresponding LLM and produces a candidate solution for the given problem.
\textbf{3) Solution Submission.} The user submits the solution by calling the \url{/api/submission} Post method. Upon receiving the submission, \texttt{CodeArena} immediately returns a \texttt{submission\_id} to the user for tracking the submission status.
\textbf{4) Isolated Execution.} Subsequently, the submitted solution is executed against predefined test cases within an isolated sandbox.
\textbf{5) Solution Persistence.} The results of the solution execution, including whether it passed or failed each test case along with any associated performance metrics, are saved in the data layer.
\textbf{6) Dynamic Evaluation.} The dynamic evaluation layer processes the execution results and updates the \texttt{dynamic points} for the submission.
\textbf{7) Submission Status.} The user can query the status of the submission with \texttt{submission\_id}. 
Detailed API usage instructions can be found in the documentation on our website.

\section{Results and Discussion}
\paragraph{Benchmarks.}
To initialize the platform, we imported APPS~\cite{hendrycksapps2021} and Mercury~\cite{du2024mercury} benchmarks to evaluate each \texttt{Code Generator}~(LLMs listed in Table~\ref{tab:benchmark}). 
Notably, \texttt{CodeArena} has sufficient flexibility to accommodate arbitrary LLM code generation benchmarks and offers online distribution and evaluation. 

\begin{table}[t]
    \caption{Leaderboard shows the code generation performance of leading open-source~($\clubsuit$) and closed-source~($\diamondsuit$) LLMs as of \textit{July 30, 2024}. \emph{DP} stands for \emph{Dynamic Points}, and the \emph{Pass} score reports the percentage of solved problems out of total problems.}
    \label{tab:benchmark}
    \resizebox{\linewidth}{!}{
        \begin{tabular}{clrr}
        \toprule
        \multicolumn{1}{c}{\textbf{Rank}} & \multicolumn{1}{c}{\textbf{Model Name}} & \multicolumn{1}{c}{\textbf{DP}} & \multicolumn{1}{c}{\textbf{Pass}} \\
        \midrule
        1                                 & $\diamondsuit$ \textbf{DeepSeek-Coder}~\cite{zhu2024deepseek}                         & 249.28                                      & 90.63\%                             \\
        2                                 & $\diamondsuit$ \textbf{GPT-4o}~\cite{achiam2023gpt}                                  & 247.32                                      & 89.06\%                             \\
        3                                 & $\diamondsuit$ \textbf{Claude-3-5-sonnet}~\cite{anthropic2024claude}                 & 227.87                                      & 74.22\%                             \\
        4                                 & $\diamondsuit$ \textbf{Gemini-1.5-flash}~\cite{team2023gemini}                       & 225.67                                      & 73.05\%                             \\
        5                                 & $\clubsuit$    \textbf{DeepSeek-Coder-V2-Lite}~\cite{zhu2024deepseek}                & 223.67                                      & 71.24\%                             \\
        6                                 & $\diamondsuit$ \textbf{Claude-3-Opus}~\cite{anthropic2024claude}                     & 221.93                                      & 69.92\%                             \\
        7                                 & $\diamondsuit$ \textbf{Gemini-1.5-pro}~\cite{team2023gemini}                         & 209.16                                      & 61.72\%                             \\
        8                                 & $\clubsuit$ \textbf{Llama-3.1-8B}~\cite{touvron2023llama}                            & 177.34                                      & 46.09\%                             \\
        9                                 & $\clubsuit$ \textbf{Llama-3-8B}~\cite{touvron2023llama}                              & 164.51                                      & 40.63\%                             \\
        10                                 & $\diamondsuit$ \textbf{GPT-4-Turbo}~\cite{achiam2023gpt}                             & 160.55                                      & 34.38\%                             \\
        11                                & $\diamondsuit$ \textbf{GPT-3.5-Turbo}~\cite{achiam2023gpt}                           & 157.70                                      & 33.98\%                             \\
        12                                & $\clubsuit$ \textbf{Mistral-Nemo}~\cite{jiang2023mistral}                            & 141.78                                      & 29.30\%                             \\
        13                                & $\clubsuit$ \textbf{CodeLlama-13b}~\cite{roziere2023code}                            & 123.15                                      & 25.39\%                             \\
        14                                & $\diamondsuit$ \textbf{Claude-3-Haiku}~\cite{anthropic2024claude}                    & 100.37                                      & 18.75\%                             \\
        15                                & $\clubsuit$ \textbf{Mistral-7B-v0.3}~\cite{jiang2023mistral}                         & 77.43                                       & 14.84\%                             \\
        16                                & $\clubsuit$ \textbf{Codestral-22B-v0.1}~\cite{jiang2023mistral}\footnotemark         & --                                          & --                                  \\
        17                                & $\diamondsuit$ \textbf{Claude-3-sonnet}~\cite{anthropic2024claude}                   & 56.17                                       & 8.98\%                              \\
        18                                & $\clubsuit$ \textbf{CodeLlama-34b}~\cite{roziere2023code}                            & 53.83                                       & 8.98\%                              \\
        19                                & $\clubsuit$ \textbf{CodeLlama-7b}~\cite{roziere2023code}                             & 50.38                                       & 6.25\%                              \\
        \bottomrule
        \end{tabular}
    }
\end{table}
\footnotetext{The scores in this snapshot were found to be erroneous and could not be re-verified, so they are marked as unavailable.}

\paragraph{Model Performance.}
In the \texttt{CodeArena} formal leaderboard, each LLM Code Generator is allowed a single attempt per problem, ensuring that dynamic point rankings are not skewed by excessive or irresponsible submissions. For demonstration purposes, we pre-registered Code Generators for several prominent LLMs and submitted their generated solutions to \texttt{CodeArena}. Detailed model inference settings are provided in Appendix~\ref{sec:model_inference}.
As shown in Table~\ref{tab:benchmark}, most closed-source LLMs significantly outperform their open-source counterparts. However, open-source LLMs do not consistently demonstrate improved performance with larger parameter scales. Notably, ``DeepSeek-Coder-V2-Lite'' achieves the highest \texttt{Pass} performance among open-source LLMs despite its relatively smaller model parameter scale.

\paragraph{Dynamic Point Changes.}
We analyze the changes in Dynamic Points ($\mathcal{DP}$) of prominent open-source ($\clubsuit$) and closed-source ($\diamondsuit$) LLMs across checkpoints ($\mathcal{CP}$) from July 30 to November 30, 2024. Compared to closed-source LLMs, open-source LLMs exhibit a clear downward trend in DP scores over time checkpoints, with ``DeepSeek-Coder-V2-Lite'' experiencing the most significant decline. In contrast, closed-source LLMs maintain stable DP scores throughout the evaluation period, even showing some improvement in the final checkpoint. 
To mitigate data contamination, LiveCodeBench~\cite{jain2024livecodebench} introduces new problems to reduce model memorization. While \texttt{CodeArena} dynamically adjusts problem weights to minimize the impact of widely leaked or trivial problems on model rankings.

\begin{figure}[t]
    \centering
    \begin{tikzpicture}
        \begin{axis}[
            width=\linewidth,
            height=0.7\linewidth,
            xlabel style={at={(1.1, -0.07)}, anchor=east, rotate=0, font=\small},
            ylabel style={at={(-0.06, 1.05)}, anchor=north, rotate=270, font=\small},
            xlabel={$\mathcal{CP}$},
            ylabel={$\mathcal{DP}$},
            xtick={1, 2, 3, 4, 5},
            xticklabels={07/30, 08/30, 09/30, 10/30, 11/30},
            ytick={140, 160, 180, 200, 220, 240}, 
            ymin=140, 
            ymax=260, 
            legend style={at={(0.5,-0.2)}, anchor=north, font=\tiny, cells = {align = left, anchor=west}},
            legend columns=3,
            grid=both,
            grid style={line width=.1pt, draw=gray!20},
            major grid style={line width=.2pt,draw=gray!50},
            minor tick num=1,
            tick label style={font=\small},
            cycle list name=color list,
        ]
            \addplot[color=Orange] coordinates {(1, 249.28) (2, 245.40) (3, 240.02) (4, 238.78) (5, 240.39)};
            \addlegendentry{$\diamondsuit$~DeepSeek-Coder}

            \addplot[color=Goldenrod] coordinates {(1, 247.32) (2, 247.41) (3, 249.63) (4, 251.64) (5, 250.59)};
            \addlegendentry{$\diamondsuit$~GPT-4o}

            \addplot[color=Mulberry] coordinates {(1, 227.87) (2, 223.74) (3, 231.40) (4, 215.70) (5, 235.09)};
            \addlegendentry{$\diamondsuit$~Claude-3-5-sonnet}

            \addplot[color=Lavender] coordinates {(1, 221.93) (2, 228.77) (3, 230.12) (4, 210.25) (5, 228.07)};
            \addlegendentry{$\diamondsuit$~Claude-3-Opus}

            \addplot[color=Cyan] coordinates {(1, 209.16) (2, 224.68) (3, 226.56) (4, 220.15) (5, 224.47)};
            \addlegendentry{$\diamondsuit$~Gemini-1.5-pro}

            \addplot[color=NavyBlue] coordinates {(1, 225.67) (2, 207.42) (3, 200.56) (4, 196.47) (5, 220.68)};
            \addlegendentry{$\diamondsuit$~Gemini-1.5-flash}

            \addplot[color=OliveGreen] coordinates {(1, 177.34) (2, 173.47) (3, 166.37) (4, 158.54) (5, 146.10)};
            \addlegendentry{$\clubsuit$~Llama-3.1-8B}

            \addplot[color=LimeGreen] coordinates {(1, 164.51) (2, 154.50) (3, 150.81) (4, 145.21) (5, 142.58)};
            \addlegendentry{$\clubsuit$~Llama-3-8B}

            \addplot[color=Apricot] coordinates {(1, 223.67) (2, 231.47) (3, 222.95) (4, 207.37) (5, 185.48)};
            \addlegendentry{$\clubsuit$~DeepSeek-Coder-V2-Lite}
        \end{axis}
    \end{tikzpicture}
    \caption{
        We trace Dynamic Point~($\mathcal{DP}$) changes of prominent open-source~($\clubsuit$) and closed-source($\diamondsuit$) LLMs over checkpoint~($\mathcal{CP}$) from 30 July to 30 Nov, 2024.
    }
    \label{fig:checkpoints}
\end{figure}
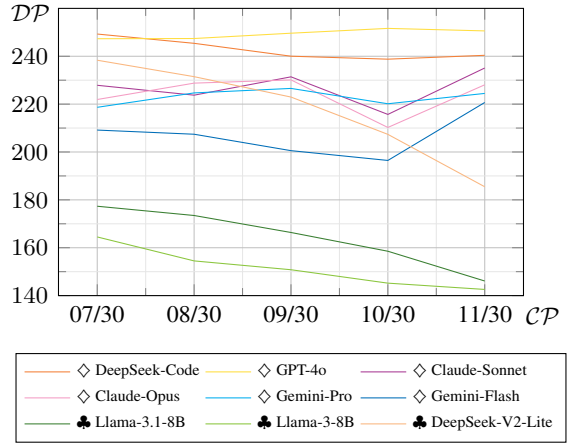

\section{System Scalability}
CodeArena is engineered for continuous operation and robust scalability. Its code execution sandbox is designed for cloud deployment, allowing for elastic expansion to accommodate fluctuating traffic demands. To ensure reliable and consistent code efficiency measurements---which can be influenced by various hardware or software factors---we currently host a cluster of sandboxes on the Google Cloud Platform~(GCP). These sandboxes all share an identical configuration~(\textit{n2-highcpu-96}) to maintain uniformity in the evaluation environment.

\section{Conclusion}
In this paper, we have introduced \texttt{CodeArena}, an online dynamic evaluation platform for LLM code generation. By integrating fresh problems, \texttt{CodeArena} maintains a challenging problem set and mitigates benchmark contamination.
Additionally, our platform provides automation-friendly APIs to facilitate user interaction and data distribution. We hope that \texttt{CodeArena} would be beneficial for creating a community-driven platform for evaluating and advancing LLM code generation.

\section*{Limitations}
While \texttt{CodeArena} significantly advances the evaluation of LLM code generation, it has limitations. It relies on external data sources like LeetCode and CodeForces, leading to issues with availability and inconsistent problem quality. Additionally, the evaluation quality depends on test cases generated by automated tools like GPT-4~\cite{achiam2023gpt}, which may not always produce exhaustive test cases. In summary, \texttt{CodeArena} is a major step forward, but it requires ongoing refinements to address these limitations.

\section*{Ethics Statement}
\paragraph{Data Management and Copyright.}
The \texttt{CodeArena} platform upholds the highest standards in data management and copyright compliance. To ensure ethical \emph{fair use}, we strictly adhere to copyright laws by using only original problems or those for which we have obtained the necessary permissions from their respective authors, ensuring they are not used for commercial purposes. We encourage researchers to utilize the platform and respect the intellectual property rights associated with all provided materials.

\paragraph{Fairness Evaluation.} 
Ensuring fairness in the evaluation of LLM-generated code is a core principle of \texttt{CodeArena}. We employ a unified prompt to invoke both open-source and closed-source LLMs within a standardized environment to avoid inconsistencies in the evaluation process. Additionally, \texttt{CodeArena} maintains an open data policy where all solutions and test cases are publicly accessible. This transparency allows the research community to scrutinize and enhance evaluation methodologies, ensuring ongoing fairness and objectivity in the benchmarking process.

\section*{Acknowledgment}
This research is supported by DSO grant DSOCL23216.
This research is also supported by A*STAR, CISCO Systems (USA) Pte. Ltd and National University of Singapore under its Cisco-NUS Accelerated Digital Economy Corporate Laboratory (Award I21001E0002). 

\newpage
\bibliography{main}

\clearpage
\appendix

\section{Proprietary Model List}
\label{sec:closed_source_models}

For closed-source LLMs, we utilize the respective provided APIs as shown in Table~\ref{tab:closed_source_models}.

\begin{table}[ht]
    \centering
    \caption{Closed-source models and their provider websites}
    \label{tab:closed_source_models}
    \resizebox{\linewidth}{!}{
        \begin{tabular}{ll}
            \toprule
            \textbf{Model Name} & \textbf{Website} \\
            \midrule
            \textbf{DeepSeek-Coder} & \url{https://chat.deepseek.com} \\
            \textbf{GPT-4o} & \url{https://chatgpt.com} \\
            \textbf{Claude-3-5-sonnet} & \url{https://www.anthropic.com} \\
            \textbf{Gemini-1.5-flash} & \url{https://gemini.google.com} \\
            \textbf{Claude-3-Opus} & \url{https://www.anthropic.com} \\
            \textbf{Gemini-1.5-pro} & \url{https://gemini.google.com} \\
            \textbf{GPT-4-Turbo} & \url{https://chatgpt.com} \\
            \textbf{GPT-3.5-Turbo} & \url{https://chatgpt.com} \\
            \textbf{Claude-3-Haiku} & \url{https://www.anthropic.com} \\
            \textbf{Claude-3-sonnet} & \url{https://www.anthropic.com} \\
            \bottomrule
        \end{tabular}
    }
\end{table}

\section{Prompt Template}
\label{sec:prompt_template}
To ensure a fair evaluation across all LLMs, we devise a unified prompt for both open-source and closed-source LLMs. This consists of two components: a \emph{system prompt} and an \emph{inference prompt}.

\begin{tcolorbox}[title=System Prompt]
\textit{You are a coding expert. You response in Pure Python code only (explicitly import all libraries). Consider each input is a string, so use \textbf{eval} to parse these inputs, and use \textbf{*} to decouple arguments.}
\end{tcolorbox}

\begin{tcolorbox}[title=Inference Prompt]
\textit{Example:}

\textit{\quad \{Example Problem Description\}}

\textit{\quad \{Example Solution\}}

\vspace{0.5em}
\textit{Given the example coding style, write the solution for the following problem. Please ONLY generate the code solution (explicitly import all libraries).}

\vspace{0.5em}
\textit{\{Problem\}}
\end{tcolorbox}

The \emph{system prompt} establishes the general instructions that guide LLMs to generate code solutions. The \emph{inference prompt} is a one-shot template with placeholders. Here, the \texttt{Example Problem Description} serves as a placeholder for the example problem statement, while the \texttt{Example Solution} provides an example solution. The \texttt{Problem} placeholder represents the actual problem that needs to be solved by the LLM.

By maintaining a uniform prompt structure, we minimize the variability introduced by different interpretation styles of LLMs. This standardized approach ensures that each LLM is assessed on an equal footing, facilitating a fair comparison of their coding capabilities. 

\section{Model Inference}
\label{sec:model_inference}
For closed-source LLMs, we utilize the respective provided APIs~(see Appendix~\ref{sec:closed_source_models}). For open-source LLMs available on HuggingFace\footnote{\url{https://huggingface.co/}}, we employ the \texttt{`text-generation'} pipeline with a temperature of~$0.7$. To achieve a balance between inference efficiency and precision, we specifically use models formatted in \texttt{`bfloat16'}. All model inferences are conducted locally on 8 NVIDIA A100 GPUs.

\section{User Groups}
\label{sec:user_groups}
In \texttt{CodeArena}, users are categorized into three distinct groups, each granted specific API permissions: \texttt{Benchmark Curators}, \texttt{Code Generators}, and \texttt{Data Readers}.

\textbf{Benchmark Curators} are pivotal in maintaining the quality of the problem repository. They are tasked with creating, refining, and expanding the set of problems available on the platform. This role involves both developing new problems and curating comprehensive test cases to ensure the problems are sufficiently challenging and evaluative. In the current configuration, the administrator fulfills the role of a benchmark curator. 

\textbf{Code Generators} can be either code-generation LLMs or human programmers. To maintain fairness and distinguish between these sub-groups, \texttt{CodeArena} registers a dedicated account for selected code generation LLMs. Each LLM user is allowed a single attempt to solve each problem. In contrast, human users are granted unlimited attempts to solve problems, and all their solutions are stored in the data repository.

\textbf{Data Readers} encompass all users interested in accessing the solution repository on the platform. These users are granted to retrieve all solution data, which is invaluable for conducting model performance analysis. To facilitate exploration, we provide a trial account (Account: \textbf{Test} / Password: \textbf{Haveatry!}) for anyone interested in browsing our data.

\begin{figure*}[ht]
    \centering
    \includegraphics[width=0.9\textwidth]{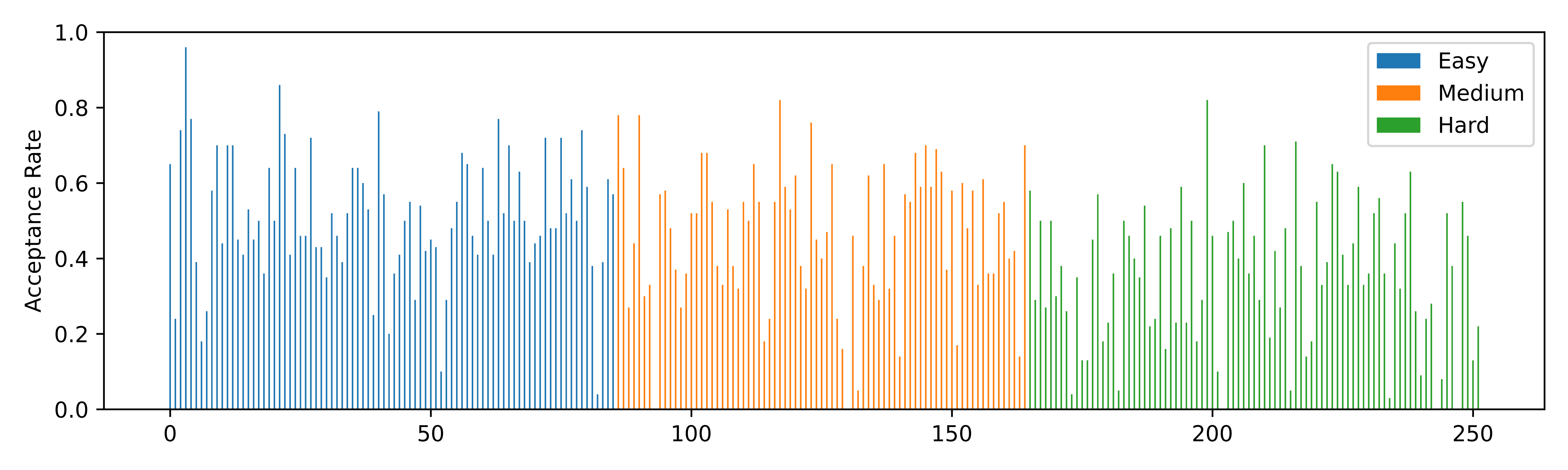}
    \caption{Acceptance Rate~(\texttt{AC}) distribution of problems clustered by the original difficulty levels inherited from Leetcode~\cite{leetcode2024}. The X-axis represents individual problems grouped by their difficulty levels, while the Y-axis indicates the AC of each problem. AC does not exhibit clear differentiation across difficulty levels.}
    \label{fig:difficulty_distribution}
\end{figure*}

\end{document}